# LIMITATIONS OF THE METHOD OF MULTIPLE-TIME-SCALES

Peter B. Kahn
Department of Physics and Astronomy
State University of New York
Stony Brook, NY 11794

and

Yair Zarmi
Department for Energy & Environmental Physics
The Jacob Blaustein Institute for Desert Research
and Physics Department
Ben-Gurion University of the Negev
Sede-Boqer Campus, 84990, Israel

ABSTRACT

In the Method of Multiple-Time-Scales (MMTS), the introduction of independent time scales and the elimination of secular terms in the fast time variable, $T_0 = t$, lead to the well-known solvability conditions. Starting from first order, "free" terms (solutions of the unperturbed equations) emerge in every order in the expansion of the approximate solution. In orders higher than first, the amplitudes of these free terms appear in the solvability conditions. Contrary to the common belief, in the MMTS analysis, these "free" terms play a role above and beyond the satisfaction of initial conditions: They make feasible mutual consistency among solvability conditions that arise in different orders. In general, this consistency may not be ensured if the "free" terms are chosen arbitrarily (e.g., set to zero, as is commonly done in many applications). If consistency is not ensured, the analysis may lead to wrong results, or allow only trivial solutions.

Limitations on the "free" terms, owing to consistency constraints among solvability conditions, can be traced to relations among resonant terms that occur in the expansion. Alternatively phrased, these limitations may occur when the Lie-Algebra of the polynomial symmetries of the linear, unperturbed part of the systems of ODE's is non-commutative. An arbitrary choice of the "free" terms can be made in the limited class of dynamical systems, for which only the commutative sub-algebra of this Lie-Algebra is "sampled" by the perturbation scheme (as in the case of a *single* harmonic oscillator with an energy conserving nonlinear perturbation).

The solvability conditions constitute a system of PDE's for the dependence of the amplitudes that appear in the expansion on the slow time scales. However, whenever the "free" amplitudes must be included to ensure a consistent expansion, these PDE's cannot determine the dependence of the solution on slow time variables beyond the first one, $T_1 = \varepsilon\, t$. The dependence on slower time scales, $T_n = \varepsilon^n\, t, n \geq 2$, must be imposed either through initial data at, say, $T_1 = 0$, or through requirements on the structure of the approximate solution (based, for example, on physical intuition) that are not related to the validity of the perturbative scheme.

These claims are illustrated through several simple examples, and then discussed in the general case.

# 1. Introduction

In the Method of Multiple Time Scales (MMTS) [1–5], the single time variable is replaced by an infinite sequence of independent time scales. The price paid for this freedom is that the well-known solvability conditions, which guarantee the elimination of secular terms in the fast variable, $T_0 = t$, impose consistency constraints on the structure of the approximate solution. Specifically, mutual consistency of solvability conditions that pertain to different orders must be ensured. This requirement limits the freedom of choice of amplitudes of "free" resonant terms, which appear in each order of the expansion. If the constraints are not obeyed, then the analysis may lead to wrong results, or allow only trivial solutions. The major consequence of this limitation is that, except for a limited class of dynamical systems (typified by the *single* harmonic oscillator with an energy conserving nonlinear perturbation), the method can only determine the dependence of the solution on the first slow variable, $T_1 = \varepsilon t$. The dependence on the higher time scales, $T_n = \varepsilon^n t, n \geq 2$, cannot be determined by the method and needs to be introduced by either selecting appropriate initial data at, say, $T_1 = 0$, or by imposing additional requirements that are based on physical or aesthetical intuition, and are unrelated to the issue of the validity of the approximation scheme.

Difficulties regarding the choice of free amplitudes in higher-order applications of the MMTS, have been noted in [6] and [7]. In [6], a weakly nonlinear harmonic oscillator with periodic forcing is analyzed. It is pointed out that, different choices of the free amplitudes, lead to conflicting results. In [7], the effect of multiple resonances on the dynamics of a suspended cable is studied. It is found that choices of the free amplitudes that do not take into account the dependence of the latter on the slow time scale, do not lead to a Hamiltonian structure of the reconstituted equation for the amplitude of the zero-order term. Such difficulties are resolved if one uses "free" amplitudes which guarantee that solvability conditions that pertain to different orders are mutually consistent. The more fundamental aspects of the consistency problem have been first raised in [8], where the connection with the symmetry structure of resonant terms in the Normal Form (NF) expansion [9] is made.

The limitations of the MMTS in higher-orders are demonstrated through the analysis of oscillatory systems of one- (Section 2), and two- (Section 3) degrees of freedom. These examples have been analyzed in the literature many times, but without attention paid to the consistency issue, which arises once orders beyond first are involved. The connection with the structure of the Lie-Algebra of the polynomial symmetries of the linear, unperturbed, part of a system of ODE's is presented in Section 4. The inability to determine, in the general case, the dependence of the amplitudes on the higher time scales, $T_n = \varepsilon^n t, n \geq 2$, is discussed in Section 5.

## 2. Systems with a single degree of freedom

### 2.1 Harmonic oscillators with cubic perturbation

The limitations of the MMTS are first shown through a comparative analysis of two simple systems, which capture the problem in its entirety: The Duffing oscillator (Hamiltonian case) and the harmonic oscillator with cubic damping (dissipative case). As these systems have been analyzed in the literature many times, some details of the analysis are omitted. The equation for the Duffing oscillator is:

$$\ddot{x} + x + \varepsilon x^3 = 0 \qquad |\varepsilon| \ll 1 \qquad (1)$$

The equation for the harmonic oscillator with cubic damping is:

$$\ddot{x} + x + \varepsilon \dot{x}^3 = 0 \qquad |\varepsilon| \ll 1 \qquad (2)$$

It is convenient to use the complex variable

$$z = x + i\,\dot{x} \qquad (3)$$

Eqs. (1) and (2) become first order equations for $z$ (a star denotes complex conjugation):

$$\dot{z} = -i\,z - \tfrac{1}{8}i\varepsilon\left(z + z^*\right)^3 \qquad (4a)$$

$$\dot{z} = -i\,z + \tfrac{1}{8}\varepsilon\left(z - z^*\right)^3 \qquad (4b)$$

Eq. (4a) is the complex versions of Eq. (1), and Eq. (4b) − of Eq. (2). In the following, equation numbers with "a" and "b" correspond to the Duffing and the damped oscillators, respectively.

To apply the MMTS, one expands $z$ in a power series in $\varepsilon$ (the near identity transformation, NIT)

$$z = z_0 + \varepsilon z_1 + \varepsilon^2 z_2 + O(\varepsilon^3) \qquad (5)$$

In addition, one replaces the derivative with respect to time by:

$$\frac{d}{dt} \to D_0 + \varepsilon D_1 + \varepsilon^2 D_2 + O(\varepsilon^3) \tag{6}$$

Here $\{D_n\}$ is an infinite sequence of partial derivatives with respect to the independent time-scale, $T_n$:

$$D_n \equiv \frac{\partial}{\partial T_n} \qquad\qquad T_n \equiv \varepsilon^n t \tag{7}$$

The order-by-order analysis through $O(\varepsilon^2)$ of the two systems discussed here yields:

$$D_0 z_0 = -i\, z_0 \tag{8}$$

$$D_0 z_1 + D_1 z_0 = -i\, z_1 - \tfrac{1}{8} i \left(z_0 + z_0^*\right)^3 \tag{9a}$$

$$D_0 z_1 + D_1 z_0 = -i\, z_1 + \tfrac{1}{8} \left(z_0 - z_0^*\right)^3 \tag{9b}$$

$$D_0 z_2 + D_1 z_1 + D_2 z_0 = -\, z_2 - \tfrac{3}{8} i \left(z_0 + z_0^*\right)^2 \left(z_1 + z_1^*\right) \tag{10a}$$

$$D_0 z_2 + D_1 z_1 + D_2 z_0 = -i\, z_2 + \tfrac{3}{8} \left(z_0 - z_0^*\right)^2 \left(z_1 - z_1^*\right) \tag{10b}$$

One now solves Eqs. (8) − (10) order-by-order.

***$O(\varepsilon^0)$:*** The solution of Eq. (8) is

$$z_0 = A \exp(-i\, T_0) \tag{11}$$

where

$$A = A(T_1, T_2, \ldots) = |A(T_1, T_2, \ldots)| e^{-i \varphi_A(T_1, T_2, \ldots)} \tag{12}$$

***$O(\varepsilon^1)$:*** To avoid a secular term in $T_0$ in the $O(\varepsilon)$ correction, $z_1$, one eliminates all contributions on the r.h.s. of Eqs. (9a) and (9b) that are proportional to $\exp(-iT_0)$. This leads to the solvability conditions:

$$D_1 A = S_1^{(a)}(A, A^*) \equiv -\left(\tfrac{3}{8}\right) i\, A^2 A^* \tag{13a}$$

$$D_1 A = S_1^{(b)}(A, A^*) \equiv -\left(\tfrac{3}{8}\right) A^2 A^* \qquad (13b)$$

In the following, $S_n(A, A^*)$ will denote the pure-$(A, A^*)$ resonant terms that appear in the $n$'th order solvability condition.

We write

$$A(T_1, T_2, \ldots) = |A| e^{-i\varphi_A} \qquad (14)$$

Using Eq. (12), Eqs. (13a) and (13b) yield the $T_1$ dependence of $A$:

$$D_1(A A^*) = 0 , \qquad \varphi_A = \tfrac{3}{8}(A A^*) T_1 + \psi_A(T_2, T_3, \ldots) \qquad (15a)$$

$$D_1(A A^*) = -\tfrac{3}{4}(A A^*)^2 \quad \Rightarrow \quad A A^* = (A A^*)_0 \big/ \left(1 + \tfrac{3}{4}(A A^*)_0 T_1\right) , \qquad D_1 \varphi_A = 0 \qquad (15b)$$

Eqs. (13a) and (13b) do not yield the dependence on higher time-scales. The $O(\varepsilon^2)$-analysis will lead to the conclusion that, depending on system type, the MMTS may not be able to yield the dependence on higher time scales.

With Eqs. (13a) and (13b) obeyed, the ($T_0$–bounded) solution for $z_1$ is found from Eqs. (9a) and (9b)

$$z_1 = \tfrac{1}{16} A^3 e^{-3i T_0} - \tfrac{3}{16} A A^{*2} e^{i T_0} - \tfrac{1}{32} A^{*3} e^{3i T_0} + B e^{-i T_0} \qquad (16a)$$

$$z_1 = \tfrac{1}{16} i A^3 e^{-3i T_0} - \tfrac{3}{16} i A A^{*2} e^{i T_0} + \tfrac{1}{32} i A^{*3} e^{3i T_0} + B e^{-i T_0} \qquad (16b)$$

$B$ is the amplitude of the (free, resonant) solution of the homogeneous equation. The amplitudes $A$ and $B$ depend on the higher time scales,

**$O(\varepsilon^2)$:** The analysis through $O(\varepsilon)$ yields a non-trivial approximation to the exact solution with an $O(\varepsilon)$ error for times of $O(1/\varepsilon)$. The issue of internal consistency of the expansion procedure arises for the first time in $O(\varepsilon^2)$. Using Eqs. (16a) and (16b) for $z_1$ in Eqs. (10a) and (10b) and exploiting Eqs. (13a) and (13b), the solvability conditions, that eliminate the $T_0$–secular term in $z_2$, are obtained:

$$D_2 A + D_1 B = S_2^{(a)}(A, A^*) - \tfrac{3}{4} i A A^* B - \tfrac{3}{8} i A^2 B^* , \quad \left(S_2^{(a)}(A, A^*) = \tfrac{51}{256} i A^3 A^{*2}\right) \qquad (17a)$$

$$D_2 A + D_1 B = S_2^{(b)}(A, A^*) - \tfrac{3}{4} A A^* B - \tfrac{3}{8} A^2 B^* \, , \qquad \left( S_2^{(b)}(A, A^*) = \tfrac{27}{256} i A^3 A^{*2} \right) \qquad (17b)$$

**Direct solution – Duffing Oscillator**

Eq. (17a) and its complex conjugate equation may be regarded as linear, first-order differential equations for the $T_1$ dependence of $(B, B^*)$ with the resonant driving terms $S_2^{(a)}(A, A^*)$ and $D_2 A$. The dependence on higher time scales is provided by the integration constants. These two equations yield

$$D_1(A^* B + A B^*) + D_2(A A^*) = 0 \qquad (18a)$$

$$D_1(A^* B - A B^*) + 2i A A^* D_2 \varphi_A = -\tfrac{3}{4} i A A^* (A^* B + A B^*) + \tfrac{51}{128} i (A A^*)^3 \qquad (19a)$$

The term $D_2(A A^*)$ in Eq. (18a) is independent of $T_1$ [Eq. (15a)]. Hence, $(A B^* + A B^*)$ has the form:

$$(A^* B + A B^*) = - D_2(A A^*) T_1 + H(T_2, T_3, \ldots) \qquad (20a)$$

Consequently, Eq. (19a) generates linear and quadratic terms in $(A B^* - A^* B)$:

$$(A^* B - A B^*) = i \left\{ \begin{array}{l} \left( -2 A A^* D_2 \varphi_A + \tfrac{51}{128}(A A^*)^3 - \tfrac{3}{4} A A^* H(T_2, T_3, \ldots) \right) T_1 + \tfrac{3}{8} A A^* D_2(A A^*) T_1^2 \\ \hspace{8cm} + J(T_2, T_3, \ldots) \end{array} \right\} \qquad (21a)$$

In Eqs. (20a) and (21a), $H$ and $J$ are real valued functions.

Over the time span $t = O(1/\varepsilon)$, for which the approximation is valid, powers of $T_1$ in $B$ need not be eliminated: They do not spoil the ordering of the expansion in progressively smaller terms because

$$\varepsilon T_1^q = \varepsilon (\varepsilon t)^q \underset{t = O(1/\varepsilon)}{\approx} O(\varepsilon) \qquad (22)$$

The $T_2$ dependence of $A$ is still undetermined. To specify it, one must either provide initial data for $A$ at, say, $T_1 = 0$, or impose other requirements. A common way to fix the $T_2$ dependence of $A$, is to demand that the solution be constructed solely out of *periodic* functions. To avoid aperiodic terms, Eq. (18a) must be broken into

$$D_2(A A^*) = 0 \qquad D_1(A^* B + A B^*) = 0 \qquad (23a)$$

Hence, $|A|$ is independent also of $T_2$. Using the same reasoning, in Eq. (19a), one must require

$$\varphi_A = -\tfrac{3}{8}(AA^*)T_1 + \tfrac{51}{256}(AA^*)^2 T_2 - \int_0^{T_2} \tfrac{3}{8}(A^*B + AB^*)dT_2 + \chi_A(T_3, T_4, ...) \tag{24a}$$

$$D_1(A^*B - AB^*) = 0 \tag{25a}$$

Consequently, one has

$$A = |A|e^{-iT_0} e^{-i\tfrac{3}{8}(AA^*)T_1} e^{i\left\{\tfrac{51}{256}(AA^*)^2 T_2 - \int_0^{T_2} \tfrac{3}{8}(A^*B + AB^*)dT_2\right\}} e^{i\chi_A(T_3, T_4, ...)} \tag{26a}$$

Here $|A|$ is independent of both $T_1$ and $T_2$.

Eq. (26a) shows that, through $O(\varepsilon^2)$, the dependence of $A$ on $T_1$ and $T_2$ is factored into a product of terms, each depending on one time scale only. [Although the $T_2$ dependence of $(A^*B + AB^*)$ is still unknown.] This pattern recurs in higher orders as well, if one requires that the solvability conditions are solved solely by periodic functions. If the requirement is not imposed, then the dependence of the amplitudes on higher time scales remains at the mercy of our choice of the initial data at, say, $T_1=0$.

With the choice that the solution retains its periodic nature, Eqs. (23a), (25a) and (13a) lead to

$$D_1(A^*B) = 0 \Rightarrow D_1 B = -\tfrac{3}{8}i AA^* B \Rightarrow \begin{cases} B(T_1, T_2, ...) = |B|e^{-i\tfrac{3}{8}(AA^*)T_1} e^{i\psi_B(T_2, T_3, ...)} \\ D_1(BB^*) = 0 \end{cases} \tag{27a}$$

Thus, the $T_1$ dependence of $\varphi_B$, the phase of the $B$, is the same as that of $\varphi_A$.

In this order, the solvability conditions leave $|B|$ free. In particular, $B = 0$ is allowed: Eqs. (13a) and (17a) remain mutually consistent, because the Lie brackets of the pure $(A, A^*)$ resonant terms vanish:

$$\{S_1^{(a)}, S_2^{(a)}\} \equiv S_2^{(a)} \frac{\partial S_1^{(a)}}{\partial A} + S_2^{(a)*} \frac{\partial S_1^{(a)}}{\partial A^*} - S_1^{(a)} \frac{\partial S_2^{(a)}}{\partial A} - S_1^{(a)*} \frac{\partial S_2^{(a)}}{\partial A^*} = 0 \tag{28a}$$

(Curly brackets are used for Lie-brackets, as square brackets will be used later on for commutators.) This guarantees that the following trivial requirement is not violated when $B = 0$:

$$D_1 D_2 A = D_2 D_1 A \tag{29}$$

In summary, if $B \neq 0$ is included in the $O(\varepsilon^2)$ solvability condition, then Eqs. (13a) and (17a) cannot yield the dependence of $A$ on *both* $T_1$ and $T_2$. They do provide both when $B = 0$ is chosen.

**Direct solution – Damped Oscillator**

Using Eq. (17b), instead of Eqs. (18a) and (19a) one now obtains

$$D_1\left\{(A^* B + A B^*)/(A A^*)^2\right\} - D_2\left(1/(A A^*)\right) = 0 \tag{18b}$$

$$D_1\left\{(A^* B - A B^*)/(A A^*) + \tfrac{9}{32} i (A A^*)\right\} + 2i D_2 \varphi_A = 0 \tag{19b}$$

Using Eq. (13b), Eq. (18b) is solved by

$$(A^* B + A B^*) = \left\{D_2\left(1/(A A^*)_0\right) T_1 + H(T_2,...)\right\}(A A^*)^2 \tag{20b}$$

The $T_1$ - linear term in Eq. (20b) does not spoil the ordering in magnitude of the zero- and first-order terms, as thanks to Eq. (15b), it generates an $O(T_1^{-1})$ leading behavior in $(A^* B + A B^*)$.

So far, the $T_2$ dependence of $A$ is not fully specified. It may be fixed in the following manner. The last term in Eq. (19b), being independent of $T_1$ [see Eq. (15b)], generates in $B$ a term proportional to $T_1^{1/2}$. For the allowed time span, $t = O(1/\varepsilon)$, this term does not destroy the ordering in magnitude in the NIT. However, as a $T_1^{1/2}$ term does not agree with the physical intuition of the expected structure of the approximate solution, it makes sense to eliminate it by splitting Eq. (19b) into:

$$D_1\left\{(A^* B - A B^*)/(A A^*) + \tfrac{9}{32} i (A A^*)\right\} = 0 \qquad D_2 \varphi_A = 0 \tag{23b}$$

An analysis similar to that leading to Eq. (21a), yields from Eqs. (20b) and (23b) the solution for $B$ Here, again, $H$ and $J$ are real valued functions of the higher time scales.:

$$B = \tfrac{1}{2}\left[\left\{D_2\left(1/(A A^*)_0\right) T_1 + H(T_2,T_3,...)\right\} A A^* + i\left\{J(T_2,T_3,...) - \tfrac{9}{32} A A^*\right\}\right] A \tag{27b}$$

This result does not lead to a secular behavior, as it yields for $BB^*$ an asymptotic $O(T_1^{-1})$ behavior, identical to that of $AA^*$. Hence, there is no need to specify $D_2(1/(AA^*)_0)$ in Eq. (27b).

In the present case, $B$ cannot be chosen arbitrarily. For example, $B = 0$ is allowed only in the trivial case, $A = 0$. This can be deduced either directly from the solution, Eq. (27b), or from the fact that, with $B$ set to zero, the solvability conditions, Eqs. (13b) and (17b), become incompatible. [Eq. (29) is not obeyed.] The reason is that the Lie brackets of the pure $(A, A^*)$ resonant terms do not vanish if $A \neq 0$:

$$D_2 D_1 A - D_1 D_2 A = \left\{ S_1^{(b)}, S_2^{(b)} \right\} = \tfrac{81}{512} i A^4 A^{*3} \tag{28b}$$

Finally, owing to the inability to choose $B = 0$, unlike the case of the Duffing oscillator, the $O(\varepsilon)$ and $O(\varepsilon^2)$ solvability conditions, Eqs. (13b) and (17b), *cannot* determine the $T_2$-dependence of $A$.

**Consistency test**

In more complicated systems, solving the solvability conditions may be sufficiently hard that one may prefer the trivial requirement of Eq. (29) (and similar ones for all higher time scales), which provides an equivalent, but simpler, procedure for deriving constraints on the free amplitudes. It *does not* introduce a new independent constraint, as commutativity of derivatives of the solutions is guaranteed, if the initial conditions are appropriately smooth. For the two systems discussed here, the pairs of solvability conditions, Eqs. (13a) and (17a), or (13b) and (17b), and Eq. (29) yield constraints on $B$:

$$D_1^2 B + \tfrac{3}{2} i (AA^*) D_1 B + \tfrac{3}{4} i A^2 D_1 B^* - \tfrac{27}{64} (AA^*)^2 B + \tfrac{9}{32} (AA^*) A^2 B^* = 0 \tag{30a}$$

$$D_1^2 B + \tfrac{3}{2} (AA^*) D_1 B + \tfrac{3}{4} A^2 D_1 B^* + \tfrac{9}{64} (AA^*)^2 B + \tfrac{9}{32} (AA^*) A^2 B^* + \tfrac{81}{512} i A^4 A^{*3} = 0 \tag{30b}$$

The conclusions of a detailed analysis of Eq. (30a) (Duffing) coincide with the results of Eq. (27a): $\varphi_B$, the phase of the free amplitude, $B$, has the same $T_1$ dependence as $\varphi_{A''}$, whereas $|B|$ may be chosen arbitrarily. In the case of the damped oscillator, Eq. (30b) allows no such freedom. For example, $B = 0$ is allowed only in the trivial case, $A = 0$, in agreement with the analysis of the solution, Eq. (27b).

**Reconstituted equation**

The consistency requirement affects another idea, often employed in MMTS analyses. The $k$'th order solvability condition provides the dependence of the amplitude of the zero-order term, $A$, on the time scale, $T_k$. Reconstruction of the full dependence of $A$ on time yields the *reconstituted equation* [10]:

$$\frac{dA}{dt} = \varepsilon D_1 A + \varepsilon^2 D_2 A + O(\varepsilon^3) \tag{31}$$

Using Eqs. (13a), (13b), (17a) and (17b), one obtains for the two oscillators considered here

$$\frac{dA}{dt} = -\varepsilon \left(\tfrac{3}{8}\right) i\, A^2 A^* + \varepsilon^2 \left\{ \begin{array}{l} \tfrac{51}{256} i\, A^3 A^{*2} \\ -\tfrac{3}{4} i\, A A^* B - \tfrac{3}{8} i\, A^2 B^* - D_1 B \end{array} \right\} + O(\varepsilon^3) \tag{32a}$$

$$\frac{dA}{dt} = -\varepsilon \left(\tfrac{3}{8}\right) A^2 A^* + \varepsilon^2 \left\{ \begin{array}{l} \tfrac{27}{256} i\, A^3 A^{*2} \\ -\tfrac{3}{4} A A^* B - \tfrac{3}{8} A^2 B^* - D_1 B \end{array} \right\} + O(\varepsilon^3) \tag{32b}$$

The free amplitude, $B$, contributes to the reconstituted equation for $A$. In the case of the Duffing oscillator, one is allowed to choose $B = 0$, converting Eq. (32a) into a "pure-$A$" equation. For the damped oscillator, the "pure-$A$" reconstituted equation is invalid, as $B = 0$ is not allowed. In both cases, a reconstituted equation can be written for:

$$\tilde{A} = A + \varepsilon B + \ldots \tag{33}$$

($\tilde{A}$ is the sum of the contributions of all the amplitudes whose $T_0$ dependence is $\exp[-iT_0]$). Using Eqs. (13a), (13b), (17a) and (17b), one finds that, a "pure-$\tilde{A}$ equation" holds for both oscillators independently of the choice of the free amplitude, $B$:

$$\frac{d\tilde{A}}{dt} = -\varepsilon \left(\tfrac{3}{8}\right) i\, \tilde{A}^2 \tilde{A}^* + \varepsilon^2 \tfrac{51}{256} i\, \tilde{A}^3 \tilde{A}^{*2} + O(\varepsilon^3) \tag{34a}$$

$$\frac{d\tilde{A}}{dt} = -\varepsilon \left(\tfrac{3}{8}\right) \tilde{A}^2 \tilde{A}^* + \varepsilon^2 \tfrac{27}{256} i\, \tilde{A}^3 \tilde{A}^{*2} + O(\varepsilon^3) \tag{34b}$$

In the case of the Duffing oscillator, one may use either $A$ or $\tilde{A}$ in the reconstituted equation. For the damped oscillator, the reconstituted equation holds only for $\tilde{A}$. Thus, although the free amplitudes do not appear explicitly in the reconstituted equations their contribution is built in.

**$O(\varepsilon^3)$:** The $O(\varepsilon^2)$ correction term in the NIT, $z_2$, contains a free term with an amplitude denoted by $C$. All amplitudes depend only on the slow time variables. The third-order solvability condition involves $A, B$ and $C$. For the Duffing oscillator, the solvability condition is given by:

$$D_1 C + D_2 B + D_3 A = S_3^{(a)}(A, A^*) + \tfrac{153}{256} i A^2 A^{*2} B + \tfrac{51}{128} i A^3 A^* B^* - \tfrac{3}{8} i A^* B^2 - \tfrac{3}{4} i A B B^*$$
$$- \tfrac{3}{4} i A A^* C - \tfrac{3}{8} i A^2 C^* \tag{35a}$$
$$\left(S_3^{(a)}(A, A^*) \equiv -\tfrac{1419}{8192} i A^4 A^{*3}\right)$$

The solvability condition in the case of the oscillator with cubic damping is:

$$D_1 C + D_2 B + D_3 A = S_3^{(b)}(A, A^*) + \tfrac{81}{256} i A^2 A^{*2} B + \tfrac{27}{128} i A^3 A^* B^* - \tfrac{3}{8} A^* B^2 - \tfrac{3}{4} A B B^*$$
$$- \tfrac{3}{4} A A^* C - \tfrac{3}{8} A^2 C^* \tag{35b}$$
$$\left(S_3^{(b)}(A, A^*) \equiv -\tfrac{567}{8192} A^4 A^{*3}\right)$$

The structure of Eqs. (35a) and (35b) is generic to all orders, $n \geq 2$. The terms with known $T_1$ dependence are: A pure $(A, A^*)$ resonant term $(S_3)$, and a combination of terms that are at least linear in each of the free amplitudes encountered in previous orders [$B$, and $B^*$ in Eqs. (35a) and (35b)]. The last term in either equation is linear in the amplitudes that appear for the first time in the solvability condition of the order considered [$C$ and $C^*$ in $O(\varepsilon^3)$]. Thus, just like Eqs. (17a) and (17b), Eqs. (35a) and (35b), and their complex conjugate equations may be regarded as first-order linear differential equations for the $T_1$ dependence of $C$ and $C^*$. The dependence on the higher time scales is accounted for by integration constants.

The choice B = C = 0 is allowed in Eq. (35a) (Duffing), because Eqs. (13a), (17a) and (34a) remain mutually consistent. [Lie brackets of each pair of pure $(A, A^*)$ resonant terms, $S_k(A, A^*)$, $k = 1, 2, 3$, vanish.] On the other hand, this choice is not allowed in Eq. (35b) (damped oscillator), because Eqs. (13b), (17b) and (34b) are then mutually inconsistent. [Lie-brackets of resonant terms do not vanish.]

The "free" amplitudes are not really free. Moreover, their dependence on higher time scales must be specified by additional requirements. This pattern recurs in higher orders.

**Connection with Lie-brackets**

The preceding discussion may be summarized in a manner that will lead to the connection with the normal form expansion, to be discussed later on. Setting the free amplitude, $B$, to zero, mutual consistency of the first-and second-order solvability conditions, through Eq. (29), may be re-written as

$$0 = D_2 D_1 A - D_1 D_2 A = D_2 S_1(A, A^*) - D_1 S_2(A, A^*) = \{S_1(A, A^*), S_2(A, A^*)\} \tag{36}$$

In Eq. (36), $\{S_1, S_2\}$ denotes the Lie brackets of $S_1$ with $S_2$. The Lie brackets of two $N$-dimensional vector-valued functions that depend on a vector variable is a vector-valued function defined as

$$\{\mathbf{F}(\mathbf{y}), \mathbf{G}(\mathbf{y})\}_i = \sum_{j=1}^{N} \left( G_j \frac{\partial F_i}{\partial y_j} - F_j \frac{\partial G_i}{\partial y_j} \right) \qquad 1 \leq i \leq N \tag{37}$$

In the case of harmonic oscillators, the $O(\varepsilon)$ and $O(\varepsilon^2)$ vector-valued functions of resonant terms:

$$\begin{pmatrix} S_1(A, A^*) \\ (S_1(A, A^*))^* \end{pmatrix} \qquad \begin{pmatrix} S_2(A, A^*) \\ (S_2(A, A^*))^* \end{pmatrix}$$

It is customary to write the Lie-brackets of these two vector-valued functions as

$$\{S_1(A, A^*), S_2(A, A^*)\}$$

The consistency requirement, Eq. (29) can, therefore, be written in the following the generic form:

$$A \text{ term linear in } B \text{ and } B^* + \{S_1(A, A^*), S_2(A, A^*)\} = 0$$

Thus, the freedom to choose the "free" amplitude, $B$, depends on the value of the Lie brackets of the pure $(A, A^*)$ resonant terms. For the two oscillators involved, one has

$$\{S_1^{(a)}(A, A^*), S_2^{(a)}(A, A^*)\} = 0 \tag{38a}$$

$$\{S_1^{(b)}(A,A^*), S_2^{(b)}(A,A^*)\} = \tfrac{81}{512} i\, A^4 A^* \tag{38b}$$

## 2.2 The Van der Pol oscillator

In the following, we analyze the solvability conditions of the Van der Pol equation:

$$\ddot{x} + x = \varepsilon\, \dot{x}(1 - x^2) \tag{39}$$

**$O(\varepsilon^1)$:**
$$D_1 A = S_1(A, A^*) = \tfrac{1}{2} A\left(1 - \tfrac{1}{4} A A^*\right) \tag{40}$$

Eq. (40) yields

$$D_1(AA^*) = AA^*\left(1 - \tfrac{1}{4} AA^*\right) \quad \Rightarrow \quad AA^* = 4(AA^*)_0 \big/ \left[(AA^*)_0 + \left(4 - (AA^*)_0\right)\exp(-T_1)\right]$$
$$D_1 \varphi_A = 0 \tag{41}$$

$\varphi_A$ and $(AA^*)_0$ do not depend on $T_1$, but may be functions of $T_2, T_3,...$

**$O(\varepsilon^2)$:**

$$D_1 B + D_2 A = S_2(A, A^*) + \tfrac{1}{2} B - \tfrac{1}{4} A A^* B - \tfrac{1}{8} A^2 B^*$$
$$\left(S_2(A,A^*) \equiv \tfrac{1}{8} i A - \tfrac{3}{16} i A^2 A^* + \tfrac{11}{256} i A^3 A^{*2}\right) \tag{42}$$

Eq. (42) and its complex conjugate equation are linear first-order equations for the $T_1$ dependence of $B$ and $B^*$. The initial conditions at, say, $T_1 = 0$ provide the dependence on higher time scales. With the aid of Eq. (40), Eq. (42) can be reduced to the following two equations:

$$D_1\left\{\exp(T_1)(A^* B + A B^*)/(AA^*)^2\right\} = D_2\left(\exp(T_1)/(AA^*)\right) = D_2\left(1/(AA^*)_0\right) \tag{43}$$

$$D_2 \varphi_A - \tfrac{1}{16} = D_1\left\{-(A^* B - A B^*)/(2 i AA^*) + \tfrac{1}{16}\log(AA^*) - \tfrac{11}{64}(AA^*)\right\} \tag{44}$$

Eq. (43) is solved by

$$(A^* B + A B^*) = \exp(-T_1)\left\{G(T_2, T_3,...) + D_2\left(1/(AA^*)_0\right) T_1\right\}(AA^*)^2 \tag{45}$$

where *G* is a real-valued function.

The $T_1$-secular term in Eq. (45) arises owing to our lack of knowledge of the $T_2$ dependence of *A*. It causes no problems for the time span over which the perturbative expansion is valid [$t=O(1/\varepsilon)$]. However, one may wish to eliminate it so that the solution is described in terms of spiraling periodic solutions only. Elimination of this term provides part of the $T_2$ dependence of *A*:

$$D_2(A A^*)_0 = 0 \Rightarrow D_2(A A^*) = 0 \tag{46}$$

$$(A^* B + A B^*) = \exp(-T_1) G(T_2, T_3, ...)(A A^*)^2 \tag{47}$$

We now turn to Eq. (44). Its l.h.s. is independent of $T_1$ [Eq. (41)]. Hence, integration over $T_1$ yields

$$(D_2\varphi_A - \tfrac{1}{16}) T_1 = -(A^* B - A B^*)/(2i A A^*) + \tfrac{1}{16}\log(A A^*) - \tfrac{11}{64}(A A^*) + H(T_2, T_3, ...) \tag{48}$$

The l.h.s. of Eq. (48) is linear in $T_1$. As the last three terms on the r.h.s. of Eq. (48) cannot account for it, this generates a linear growth of *B* with $T_1$. This behavior does not spoil the ordering in the expansion for $t=O(1/\varepsilon)$. However, it is an unphysical behavior. To avoid it, the l.h.s. of Eq. (48) must vanish, completing the $T_2$ dependence of *A*:

$$D_2\varphi_A = \tfrac{1}{16} \Rightarrow \varphi_A = \tfrac{1}{16} T_2 + \psi_A(T_3, T_4, ...) \tag{49}$$

$$(A^* B - A B^*) = \left\{ \tfrac{1}{16}\log(A A^*) - \tfrac{11}{64}(A A^*) + H(T_2, T_3, ...) \right\} 2i(A A^*) \tag{50}$$

From Eqs. (47) and (50), we find

$$B = \left\{ \begin{array}{c} \tfrac{1}{2} G(T_2, T_3, ...)\exp(-T_1)(A A^*) + i H(T_2, T_3, ...) \\ + \tfrac{1}{16} i \log(A A^*) - \tfrac{11}{64} i (A A^*) \end{array} \right\} A \tag{51}$$

Again, the choice of *B* is restricted. For example, $B = 0$ is only possible in the trivial cases, $|A| = 0, 2$. This conclusion follows directly from the solution, Eq. (51), or from the fact that Eqs. (40) and (42) then become incompatible [the Lie brackets of the pure $(A, A^*)$ resonant terms do not vanish].

Finally, the $T_2$ dependence of $A$ has been determined by requirements that are unrelated to the issue of the validity of the perturbative expansion.

## $O(\varepsilon^3)$

$$\begin{aligned}
&D_1 C + D_2 B + D_3 A = S_3(A, A^{*+}) + \\
&\tfrac{1}{8} iB - \tfrac{3}{8} i A A^* B + \tfrac{33}{256} i (A A^*)^2 B - \tfrac{3}{16} i A^2 B^* + \tfrac{11}{128} i A^3 A^* B^* - \tfrac{1}{8} A^* B^2 - \tfrac{1}{4} A B B^* + \\
&\tfrac{1}{2} C - \tfrac{1}{4} A A^* C - \tfrac{1}{8} A^2 C^* \\
&\qquad\qquad\qquad \left( S_3(A, A^*) \equiv -\tfrac{3}{128} A^2 A^* + \tfrac{11}{1024} A^3 A^{*2} - \tfrac{13}{8192} A^4 A^{*3} \right)
\end{aligned} \qquad (52)$$

The generic structure of the solvability conditions emerges again; hence the arguments discussed previously apply here as well.

## 3. Systems with two degrees of freedom

In this Section we demonstrate the crucial role played by "free" amplitudes in avoiding trivial solutions in an MMTS analysis of systems with two degrees of freedom.

### 3.1 Hamiltonian system of two coupled harmonic oscillators

Consider a system of two coupled harmonic oscillators with equal unperturbed frequencies ($\omega_1=\omega_2=1$) with Hamiltonian:

$$H = \tfrac{1}{2}\dot{x}_1^{\,2} + \tfrac{1}{2}x_1^{\,2} + \tfrac{1}{2}\dot{x}_2^{\,2} + \tfrac{1}{2}x_2^{\,2} + \tfrac{1}{2}\varepsilon x_1^{\,2} x_2^{\,2} \qquad (53)$$

Again, we use the complex notation

$$z_i = x_i + i\,\dot{x}_i \qquad i = 1, 2 \qquad (54)$$

The equation of motion will be written in explicitly for oscillator no.1. [The equation for oscillator no.2 is obtained through the transformation of indices: $(1,2) \to (2,1)$.]

$$\dot{z}_1 = -i\,z_1 - \tfrac{1}{8} i\,\varepsilon(z_1 + \bar{z}_1)(z_2 + \bar{z}_2)^2 \qquad (55)$$

One now expands $z_i$ in perturbation series, and writes for the zero-order approximation

$$z_{i,0} = A_i e^{-i T_0} \qquad A_i = |A_i| e^{-i \varphi_i} \tag{56}$$

Applying the formalism of the MMTS, one derives the following solvability conditions.

**$O(\varepsilon^1)$**
$$D_1 A_1 = -\tfrac{1}{4} i A_1 A_2 A_2^* - \tfrac{1}{8} i A_1^* A_2^{\,2} \tag{57}$$

Eq. (57) generates a $T_1$ dependence in the amplitudes and the phases:

$$D_1(A_1 A_1^*) = -D_1(A_2 A_2^*) = \tfrac{1}{8} i \left( A_1^{\,2} A_2^{*\,2} - A_1^{*\,2} A_2^{\,2} \right) \tag{58}$$

$$D_1 \varphi_1 = \tfrac{1}{4} A_2 A_2^* + \tfrac{1}{16} \frac{\left( A_1^{\,2} A_2^{*\,2} + A_1^{*\,2} A_2^{\,2} \right)}{A_1 A_1^*} \tag{59}$$

Eq. (58) describes the efficient energy-transfer between the two oscillators, a consequence of their resonating unperturbed frequencies. This is the germ of the problem. Whereas the whole system is energy conserving, each degree of freedom suffers "dissipation" as energy is transferred to the other degree of freedom. Thus, one expects to encounter the same consistency problems observed in the case of a single oscillator with a dissipative perturbation, as demonstrated in Section 2.

**$O(\varepsilon^2)$**

$$\begin{aligned}
D_1 B_1 + D_2 A_1 = & \\
\tfrac{1}{64} i A_1^{\,3} A_2^{*\,2} + \tfrac{9}{128} i A_1^{\,2} A_1^* A_2 A_2^* &+ \tfrac{3}{64} i A_1 A_1^{*\,2} A_2^{\,2} + \tfrac{9}{256} i A_1 A_2^{\,2} A_2^{*\,2} + \tfrac{1}{32} i A_1^* A_2^{\,3} A_2^* \\
- \tfrac{1}{8} i A_2^{\,2} B_1^* &- \tfrac{1}{4} i A_2 A_2^* B_1 - \tfrac{1}{4} i A_1 A_2 B_2^* - \tfrac{1}{4} i A_1 A_2^* B_2 - \tfrac{1}{4} i A_1^* A_2 B_2
\end{aligned} \tag{60}$$

In Eq. (60), one is not free to choose the amplitudes $B_i$ ($i = 1, 2$) arbitrarily. For the sake of simplicity, we do not present the general analysis, and show that the choice $B_i = 0$ is allowed only in trivial cases. Although the analysis can be carried out by directly solving the solvability conditions, we use the much simpler route of the commutativity of the derivatives with respect to different time scales. Applying the $D_2$ derivative to Eq. (58) and $D_1$ – to Eq. (60) and setting $B_i = 0$, we obtain

$$0 = D_1 D_2 A_1 - D_2 D_1 A_1 = -\tfrac{7}{1024} A_1^* \left( A_1^{\,4} A_2^{*\,2} - 2 A_1^{\,2} A_1^{*\,2} A_2^{\,2} + A_2^{\,4} A_2^{*\,2} \right) \tag{61}$$

Eq. (61) and the corresponding equation for $A_2$ are obeyed only in one of the following cases:

$$A_1 = 0, \qquad A_2 = 0, \qquad A_2 = \pm A_1 \tag{62}$$

In the first and second cases, one of the oscillators is not excited. In the third, the two oscillators oscillate with identical amplitudes, so that, effectively, they are uncoupled. Thus, a choice of the "free" amplitudes that disregards their role in ensuring consistency allows only trivial solutions.

**Connection with Lie-brackets**

As there are two degrees of freedom, the resonant terms are four-dimensional vectors given by

$$\mathbf{S}_j = \begin{pmatrix} a_j A_1^{n_{j1}} A_2^{m_{j1}} A_1^{*\, p_{j1}} A_2^{*\, q_{j1}} \\ b_j A_1^{n_{j2}} A_2^{m_{j2}} A_1^{*\, p_{j2}} A_2^{*\, q_{j2}} \\ \left(a_j A_1^{n_{j1}} A_2^{m_{j1}} A_1^{*\, p_{j1}} A_2^{*\, q_{j1}}\right)^* \\ \left(b_j A_1^{n_{j2}} A_2^{m_{j2}} A_1^{*\, p_{j2}} A_2^{*\, q_{j2}}\right)^* \end{pmatrix} \tag{63}$$

(The components in the vectors are ordered from top to bottom in the order $i = z_{1,0}, z_{2,0}, z_{1,0}^*$ and $z_{2,0}^*$.) Each monomial obeys the resonance condition

$$n_{ji} + m_{ji} - p_{ji} - q_{ji} = 1 \tag{64}$$

The Lie-bracket of any two resonant terms, $\mathbf{S}_1$ and $\mathbf{S}_2$, is also a four-dimensional vector. For example, the $z_{1,0}$-component of the Lie-bracket is:

$$\begin{aligned}\{\mathbf{S}_1,\mathbf{S}_2\}^{z_{1,0}} &= S_2^{z_{1,0}} \frac{\partial S_1^{z_{1,0}}}{\partial z_{1,0}} + S_2^{z_{1,0}*} \frac{\partial S_1^{z_{1,0}}}{\partial z_0^*} + S_2^{z_{2,0}} \frac{\partial S_1^{z_{1,0}}}{\partial z_{2,0}} + S_2^{z_{2,0}*} \frac{\partial S_1^{z_{1,0}}}{\partial z_{2,0}^*} \\ &\quad - S_1^{z_{1,0}} \frac{\partial S_2^{z_{1,0}}}{\partial z_{1,0}} - S_1^{z_{1,0}*} \frac{\partial S_2^{z_{1,0}}}{\partial z_{1,0}^*} - S_1^{z_{2,0}} \frac{\partial S_2^{z_{1,0}}}{\partial z_{2,0}} - S_1^{z_{2,0}*} \frac{\partial S_2^{z_{1,0}}}{\partial z_{2,0}^*}\end{aligned} \tag{65}$$

A similar expression holds for the $z_{2,0}$-component of the Lie-brackets. The $z_{1,0}^*$ and $z_{2,0}^*$ components are the complex conjugates of the first two.

In general, the Lie brackets, Eq. (65), of two resonant vectors that obey Eq. (64), do not vanish, even if the coefficients $a_i$ and $b_i$ are pure imaginary. The Lie brackets do vanish, only in the trivial case, when one of the two vectors, say $\mathbf{S}_1$ is just the unperturbed term of the dynamical equation, and the vanishing of the Lie brackets is nothing but an alternative definition of resonance [9]. This is the case when

$$
\begin{aligned}
n_{11} &= m_{12} = 1 \\
m_{11} &= p_{11} = q_{11} = n_{12} = p_{12} = q_{12} = 0
\end{aligned}
\tag{66}
$$

Consequently, in general, the first-and second-order solvability conditions are not mutually consistent, unless the first-order "free:" amplitudes are incorporated in a manner that allows such consistency.

### 3.2 The Mathieu equation – where the freedom is twofold

The Mathieu equation arises in many physical situations, e.g., in solid-state physics and in accelerator design. The equation is:

$$
\ddot{x} + (a + 2\varepsilon \cos t)x = 0 \qquad |\varepsilon| \ll 1 \tag{67}
$$

Eq. (67) represents a system with two degrees of freedom, as it describes the onset of instabilities in one degree of freedom, $x(t)$, owing to its weak coupling to another oscillatory degree of freedom. Hence, the consistency problems discussed in Section 3.1 ought to apply in this case as well. Consider the vicinity of the first resonance in this equation, occurring when

$$
a = \tfrac{1}{4} + \mu \qquad |\mu| \ll 1 \tag{68}
$$

From the physical point of view, $\varepsilon$ and $\mu$ are independent parameters. A NF expansion of Eq. (67) encounters no problems in treating $\varepsilon$ and $\mu$ as independent parameters. A second-order NF analysis yields for the stability domain of the solution [11]:

$$
-\varepsilon - \tfrac{1}{2}\varepsilon^2 \le \mu \le +\varepsilon - \tfrac{1}{2}\varepsilon^2 \tag{69}
$$

In a MMTS analysis of Eq. (67) that treats $\varepsilon$ and $\mu$ as independent, one defines time scales:

$$
T_{kl} \equiv \varepsilon^k \mu^l t \qquad D_{kl} \equiv \frac{\partial}{\partial T_{kl}} \qquad k,l = 0,1,2,\ldots \tag{70}
$$

Writing the zero-order solution as

$$z_{00} = A(T_{10}, T_{01},...)e^{-i T_{00}} \tag{71}$$

The first-order solvability conditions are found to be

$$D_{10}A = -\tfrac{1}{4}i A^* \qquad\qquad D_{01}A = -\tfrac{1}{2}i A \tag{72}$$

The only solution consistent with Eqs. (72) is

$$A = 0 \tag{73}$$

Consistency of the solvability conditions in the next order forces the first-order term to vanish as well. Thus, the well-known fact, that a MMTS analysis with several expansion parameters may encounter difficulties, is a consequence of the inability to ensure the mutual consistency of solvability conditions. [It is easy to show that the Lie brackets of the two resonant terms in Eqs. (72) do not vanish.] This is why it is customary to cast Eq. (67) as a single-parameter problem by writing

$$\mu = a_1 \varepsilon + a_2 \varepsilon^2 + a_3 \varepsilon^3 + O(\varepsilon^4) \tag{74}$$

We now expand

$$x = x_0 + \varepsilon x_1 + \varepsilon^2 x_2 + O(\varepsilon^3) \tag{75}$$

and write

$$x_0 = A e^{i T_0} + A^* e^{-i T_0} \qquad\qquad A \equiv A(T_1, T_2,...) \tag{76}$$

The solvability conditions can be written in a concise form by use of the Pauli $\sigma$ - matrices

**_O(e)_:**

$$D_1 \begin{pmatrix} A \\ A^* \end{pmatrix} = (i a_1 \sigma_z - \sigma_y) \begin{pmatrix} A \\ A^* \end{pmatrix} \tag{77}$$

Eq. (77) leads to

$$D_1^2 \begin{pmatrix} A \\ A^* \end{pmatrix} = -(a_1^2 - 1) \begin{pmatrix} A \\ A^* \end{pmatrix} \tag{78}$$

Stability requires

$$|a_1| > 1 \tag{79}$$

Thus, the boundary of the stability domain is at $a_1 = \pm 1$.

## $O(e^2)$:

$$D_1 \begin{pmatrix} B \\ B^* \end{pmatrix} = (i a_1 \sigma_z - \sigma_y) \begin{pmatrix} B \\ B^* \end{pmatrix} - D_2 \begin{pmatrix} A \\ A^* \end{pmatrix} + i g \sigma_z \begin{pmatrix} A \\ A^* \end{pmatrix} \tag{80}$$

$$\left( g = \left[ (1 - a_1^2) + a_2 + \tfrac{1}{2} \right] \right)$$

Eqs. (77) and (80) lead directly to

$$D_1^2 \begin{pmatrix} B \\ B^* \end{pmatrix} + (a_1^2 - 1) \begin{pmatrix} B \\ B^* \end{pmatrix} = -2(i a_1 \sigma_z - \sigma_y) D_2 \begin{pmatrix} A \\ A^* \end{pmatrix} - 2 a_1 g \begin{pmatrix} A \\ A^* \end{pmatrix} \tag{81}$$

## Consistency test

To find the implications of the requirement for mutual consistency of the solvability conditions, Eqs. (77) and (80), we start with the common choice for the free amplitude, $B = 0$. Eq. (80) then yields:

$$D_2 \begin{pmatrix} A \\ A^* \end{pmatrix} = i g \sigma_z \begin{pmatrix} A \\ A^* \end{pmatrix} \tag{82}$$

Now apply $D_2$ to Eq. (77) and $D_1$ - to Eq. (82). For the results to coincide, the following must hold:

$$g = 0 \quad \Rightarrow \quad (1 - a_1^2) + (a_2 + \tfrac{1}{2}) = 0 \tag{83}$$

At the boundary of the stability domain, Eq. (83) leads to

$$a_2 \xrightarrow[a_1^2 \to 1]{} -\tfrac{1}{2} \tag{84}$$

When $B \neq 0$, one returns to Eq. (81). To avoid $T_1$ - secular terms in $B$, so that $B$ is periodic in $T_1$, the r.h.s. of Eq. (81) must vanish, leading to

$$D_2\begin{pmatrix} A \\ A^* \end{pmatrix} = \frac{a_1 g}{a_1^2 - 1}(ia_1\sigma_z - \sigma_y)\begin{pmatrix} A \\ A^* \end{pmatrix} \quad \Rightarrow \quad D_2^2\begin{pmatrix} A \\ A^* \end{pmatrix} = -\frac{(a_1 g)^2}{a_1^2 - 1}\begin{pmatrix} A \\ A^* \end{pmatrix} \quad (85)$$

This is not required for the validity of the expansion through $t = O(1/\varepsilon)$, but enables the determination of the $T_2$ dependence of $A$, which, otherwise would be impossible. Eqs. (77) & (85) are solved by

$$\begin{pmatrix} A \\ A^* \end{pmatrix} = \exp\left[(ia_1\sigma_z - \sigma_y)T_1\right]\exp\left[\frac{a_1 g}{a_1^2 - 1}(ia_1\sigma_z - \sigma_y)T_2\right]\mathbf{A}_0(T_3,...) \quad (86)$$

Using Eq. (85), the solution for Eq. (80) is found to be

$$\begin{pmatrix} B \\ B^* \end{pmatrix} = \tfrac{1}{2}\frac{g}{a_1^2 - 1}\sigma_x\begin{pmatrix} A \\ A^* \end{pmatrix} + \exp\left[(ia_1\sigma_z - \sigma_y)T_1\right]\mathbf{B}_0(T_2, T_3,...) \quad (87)$$

To guarantee that $A$ and $B$ are bounded as $a_1^2$ tends to 1 from within the stability domain, one needs

$$g = \underset{(a_1^2 \to 1)}{O(a_1^2 - 1)} \qquad a_2 \xrightarrow[a_1^2 \to 1]{} -\tfrac{1}{2} + O(a_1^2 - 1) \quad (88)$$

Eqs. (83) and (84) represent a particular case of the more general requirement of Eq. (88).

The analysis in higher orders proceeds in the same manner. The results can be summarized as follows. If the free amplitudes are incorporated in the expansion in a manner that guarantees mutual consistency of the solvability conditions, then the results are identical to those of the NF expansion [11]: The values of the coefficients $a_2$, $a_3$, ..., are not constrained *inside* the stability domain. The MMTS only dictates their values at the boundary of the stability domain. If the free amplitudes are set to zero, then the method exploits the freedom in the coefficients $a_2$, $a_3$, ..., to guarantee that the consistency constraints are fulfilled. Consequently, $a_2$, $a_3$, ..., become related to $a_1$. Namely, the consistent analysis has to be carried out along lines within the $\mu - \varepsilon$ plane. Each line is specified by a choice of $a_1$. The whole stability domain is covered by varying $a_1$ from $-1$ to $+1$.

That a difficulty exists in the determination of the coefficients $a_1$, $a_2$, $a_3$, ... was observed in [4]. However, as the consistency of solvability conditions was not fully exploited, additional time scales with powers of $\varepsilon^{1/2}$ were introduced in order to resolve the problem.

## 4. Symmetry structure of resonant terms and consistent MMTS expansion

The fundamental aspect of the consistency problem has been first raised in [8], where the free resonant terms, are omitted in all orders of the expansion. This leads to the conclusion is that, in general, the MMTS expansion cannot be consistently extended beyond first order. The main issue in [8] is the important analysis of the algebra associated with the polynomial symmetries of the linear, unperturbed, part of a system of ODE's. We briefly review the analysis of [8].

Consider an $N$-dimensional system of coupled ODE's

$$\frac{d\mathbf{z}}{dt} = \mathbf{A}\,\mathbf{z} + \sum_{n} \varepsilon^{n}\, \mathbf{F}^{(n)}(\mathbf{z}) \tag{89}$$

In Eq. (89), $\mathbf{z}$ is an $N$-dimensional vector; $\mathbf{A}$ is an $N$x$N$ matrix, for simplicity, assumed to be diagonal:

$$\mathbf{A} = \begin{pmatrix} \lambda_1 & \cdots & 0 \\ \cdot & \cdot & \cdot \\ \cdot & & \cdot \\ 0 & \cdots & \lambda_N \end{pmatrix} \tag{90}$$

$\mathbf{F}^{(n)}$ is a vector-valued function of homogeneous polynomials of degree $n$.

The normal form expansion [9] is comprised of the Near Identity transformation (NIT)

$$\mathbf{z} = \mathbf{z}_0 + \sum_{n} \varepsilon^{n}\, \mathbf{z}_n(\mathbf{z}_0) \tag{91}$$

and the normal form, NF, (an asymptotic expansion of the dynamical equation for $\mathbf{z}_0$)

$$\frac{d\mathbf{z}_0}{dt} = \mathbf{A}\,\mathbf{z}_0 + \sum_{n} \varepsilon^{n}\, \mathbf{S}^{n}(\mathbf{z}_0) \qquad \left( \mathbf{S}^{n}(\mathbf{z}_0) \equiv \sum_{l=0}^{N-1} \mathbf{C}_{l}^{n}\, \mathbf{S}_{l}^{n}(\mathbf{z}_0) \right) \tag{92}$$

In Eq. (92), $\mathbf{C}_{l}^{n}$ is an $N$x$N$ matrix of coefficients, reflecting the nature of the nonlinear perturbation, as well as the choice of free resonant terms, which emerge in every order of the expansion. $\mathbf{S}_{l}^{n}(\mathbf{z}_0)$ ($0 \le l \le N-1$) are $N$ independent vectors, with components that are resonant monomials of degree $n$. In [8] an algorithm for the construction of $\mathbf{S}_{l}^{n}(\mathbf{z}_0)$ is presented. A monomial

$$\prod_{j=1}^{N} z_{0,j}{}^{m_j}$$

is *resonant*, if for some $1 \leq s \leq N$, a resonance condition holds among the unperturbed eigenvalues [9]:

$$\lambda_s = (\mathbf{m} \cdot \boldsymbol{\lambda}) = \sum_{j=1}^{N} m_j \lambda_j \tag{93}$$

By definition, the Lie-brackets of $\mathbf{S}_l^n(\mathbf{z}_0)$ with the unperturbed, linear part of Eq. (32) vanish:

$$\{\mathbf{A}\mathbf{z}_0, \mathbf{S}_l^n\} = 0 \tag{94}$$

However, in general, the Lie-brackets of different resonant vectors need not vanish:

$$\{\mathbf{S}_l^n, \mathbf{S}_k^m\} \neq 0 \qquad n,m \geq 1 \qquad n \neq m \tag{95}$$

(They always vanish for $k = l = 1$.) As a result, the algebra generated by the vector fields

$$X_l^n = \mathbf{S}_l^n \cdot \frac{d}{d\mathbf{z}_0} = \sum_{i=1}^{N} (\mathbf{S}_l^n)_i \frac{\partial}{\partial z_{0,i}} \qquad n \geq 1 \tag{96}$$

may be non-commutative, as one has

$$[X_l^n, X_k^m] = -\{\mathbf{S}_l^n, \mathbf{S}_k^m\} \cdot \frac{d}{d\mathbf{z}_0} \neq 0 \tag{97}$$

(Note that the sub-algebra with $k = l = 1$ is commutative.) The dimensionality of the algebra depends on the eigenvalues, $\lambda_1, \ldots, \lambda_N$. If the number of integer-vector combinations that obey the resonance condition, Eq. (93), is finite (as is the case when all eigenvalues lie in a Poincaré domain [9]), then the algebra is finite-dimensional. The algebra is infinite dimensional if the number of resonant combinations is infinite (as may happen if the eigenvalues lie in a Siegel domain [9]).

Consider the case when the free resonant terms in the higher-order terms in the NIT, Eq. (91), are set to zero in both the MMTS and NF expansions. Then, order-by-order, the resonant terms that appear in the solvability conditions in the MMTS analysis and in the NF, Eq. (92), are identical in form. The commutator of derivatives with respect to two time-scales, $T_n$ and $T_m$ ($n \neq m$), is then given by [8]

$$[D_n, D_n]\mathbf{z}_0 = D_n D_m \mathbf{z}_0 - D_m D_n \mathbf{z}_0 = -\{\mathbf{S}^n(\mathbf{z}_0), \mathbf{S}^m(\mathbf{z}_0)\} \qquad (98)$$

The commutator must vanish when applied to $\mathbf{z}_0$. This is ensured only if the corresponding Lie-brackets vanish. If the Lie brackets do not vanish, then the free amplitudes cannot be set to zero (or chosen arbitrarily). Vanishing of the commutator then constitutes a consistency constraint that the free amplitudes must obey. Eqs. (30a) and (30b) provide an example of this rule.

Consider the case of a harmonic oscillator. Denoting the zero-order approximation to the solution, $z(t)$, by $z_0$, and using the complex notation of Section 2 and the definitions in [8], the space of resonant vectors is two-dimensional, with basis vectors:

$$\mathbf{S}_0^n = \begin{pmatrix} (z_0 z_0^*)^n z_0 \\ (z_0 z_0^*)^n z_0^* \end{pmatrix} \qquad \mathbf{S}_1^n = \begin{pmatrix} (z_0 z_0^*)^n (-i) z_0 \\ (z_0 z_0^*)^n (+i) z_0^* \end{pmatrix} \qquad (99)$$

The unperturbed part of the equation is

$$\mathbf{Z}_0 = \begin{pmatrix} -i z_0 \\ i z_0^* \end{pmatrix} \qquad (100)$$

That $\mathbf{S}_l^n$ ($l = 0, 1$) are both resonant vectors is easily verified by checking that [9]

$$\{\mathbf{Z}_0, \mathbf{S}_l^n\} = 0 \qquad (101)$$

However, not all the Lie-brackets $\{\mathbf{S}_l^n . \mathbf{S}_k^m\}$ vanish in this case. As a result, the algebra of the vector fields generated by $\mathbf{S}_l^n$ is non-commutative [8]:

$$\begin{aligned} \left[X_0^n, X_0^m\right] &= 2(m-n) X_0^{n+m} \\ \left[X_0^n, X_1^m\right] &= 2m X_1^{n+m} \\ \left[X_1^n, X_1^m\right] &= 0 \end{aligned} \qquad (102)$$

Although there are only two eigenvalues in this case, $\pm i$, the algebra is infinite dimensional, because there is an infinite number of resonant combinations:

$$i = (n+1)(i) + n(-i) \quad n \geq 1$$

The coefficient matrices, $\mathbf{C}_l^n$, have the structure [8]

$$\mathbf{C}_l^n = \begin{pmatrix} a_l^n & 0 \\ 0 & (a_l^n)^* \end{pmatrix} \qquad l = 0,1 \tag{103}$$

Consequently, the normal form equation becomes

$$\frac{dz_0}{dt} = -iu + \left\{ \sum_n (a_0^n - i a_1^n)(z_0 z_0^*)^n \right\} z_0 \tag{104}$$

One may assume that $a_0^n$ and $a_1^n$ are real. Otherwise, one separates their real and imaginary parts.

If the perturbation has a dissipative component, the coefficients of the resonant terms in Eq. (104) have non-vanishing real parts. The perturbation "samples" the vectors $\mathbf{S}_0^n$, which yield non-vanishing Lie-brackets for the resonant terms. It then "samples" the non-commutative part of the algebra of Eq. (102), generated by the vector fields $X_0^n$. For a conservative perturbation, all resonant terms in Eq. (104) have pure imaginary coefficients. Then all pairs of resonant terms then have vanishing Lie-brackets: The perturbation "samples" only the vectors $\mathbf{S}_1^n$. (the commutative sub-algebra of $X_1^n$ ). To be specific, consider the oscillators analyzed in Section 2: The Duffing oscillator (equations denoted by "a") and the damped oscillator ("b"). Their normal form analysis yields [11]:

NIT

$$z = z_0 + \varepsilon \left\{ \tfrac{1}{16} z_0^3 - \tfrac{3}{16} z_0 z_0^{*2} - \tfrac{1}{32} z_0^{*3} + \alpha\, z_0^2 z_0^* \right\} + O(\varepsilon^2) \tag{105a}$$

$$z = z_0 + \varepsilon \left\{ \tfrac{1}{16} i z_0^3 - \tfrac{3}{16} i z_0 z_0^{*2} + \tfrac{1}{32} i z_0^{*3} + f_1(z_0 z_0^*) z_0 \right\} + O(\varepsilon^2) \tag{105b}$$

NF

$$\frac{dz_0}{dt} = -i z_0 - \tfrac{3}{8}\varepsilon\, i z_0^2 z_0^* + \varepsilon^2 i\left\{ \left(\tfrac{51}{256}\right) - \left(\tfrac{3}{8}\right)(\alpha + \alpha^*) \right\} z_0^3 z_0^{*2} + O(\varepsilon^3) \tag{106a}$$

$$\frac{dz_0}{dt} = -i z_0 - \tfrac{3}{8}\varepsilon\, z_0^2 z_0^* + \varepsilon^2 \left\{ \left(\tfrac{27}{256}\right) i z_0^3 z_0^{*2} - \left(\tfrac{3}{8}\right)(f_1 + f_1^*) z_0^2 z_0^* + \tfrac{3}{4}\frac{\partial f_1}{\partial (uu^*)} z_0^3 z_0^{*2} \right\} + O(\varepsilon^3) \tag{106b}$$

In the case of the Duffing oscillator, the free resonant term in the first-order correction, $z_1$, has a constant coefficient, $\alpha$, and in the case of the oscillator with cubic damping, the free resonant term is proportional to an arbitrary function $f_1(z_0 z_0^*)$. As the resonant terms in $z_1$ in both systems are indeed free in the NF expansion, one may drop them. (A normal form expansion without free resonant terms was called the *distinguished choice* [12].) Eqs. (106a) and (106b) become

$$\frac{dz_0}{dt} = -i\, z_0 - \tfrac{3}{8}\varepsilon\, i\, z_0^{\,2}\, z_0^{\,*} + \varepsilon^{\,2}\, i\left(\tfrac{51}{256}\right) z_0^{\,3} z_0^{\,*2} + O(\varepsilon^{\,3}) \tag{107a}$$

$$\frac{dz_0}{dt} = -i\, z_0 - \tfrac{3}{8}\varepsilon\, z_0^{\,2}\, z_0^{\,*} + \varepsilon^{\,2}\left(\tfrac{27}{256}\right) i\, z_0^{\,3} z_0^{\,*2} + O(\varepsilon^{\,3}) \tag{107b}$$

If in the MMTS analysis, one chooses $B = 0$, the solvability conditions, Eqs. (17a) and (17b), become

$$D_1 A = -\tfrac{3}{8} i\, A^{\,2} A^{*} \qquad D_2 A = \tfrac{51}{256} i\, A^{\,3} A^{*\,2} \tag{108a}$$

$$D_1 A = -\tfrac{3}{8} A^{\,2} A^{*} \qquad D_2 A = \tfrac{27}{256} i\, A^{\,3} A^{*\,2} \tag{108b}$$

Comparing the normal forms, Eqs. (107a) and (107b) with the solvability conditions, Eqs. (108a) and (108b), one sees that the equations in the two methods have the same structure. In the case of the Duffing oscillator, the two conditions in Eq. (108a) are consistent with one another despite the fact that the free amplitude, $B$, has been dropped. On the other hand, in Eq. (108b) (oscillator with cubic damping), they are not consistent *because* the "free" amplitude, $B$, has been dropped.

This difference results from the different the symmetry properties of Lie-brackets of resonant terms in the two cases. In the absence of free amplitudes, Eq. (29) requires that that the Lie-brackets of the first- and second-order terms in the normal form vanish. This requirement is obeyed in the case the Duffing oscillator, but not in the case of the damped oscillator. Alternatively, in the Duffing oscillator, only the commutative sub-algebra of $X_1^{\,n}$ is "sampled" by the perturbation. In the case of the damped oscillator, non-commutative parts of the algebra are also "sampled". Then, Eq. (29) cannot be satisfied if the "free" amplitude, $B$, is omitted. Inclusion of a free $B$-amplitude ensures that Eq. (29) can be satisfied, *despite the non-commutative nature of the algebra.*

The conclusion drawn from the examples discussed here for the general case is obvious. In a specific problem, if in the NF expansion, terms that pertain to two different orders, $n$ and $m$, have vanishing Lie-brackets, then in the MMTS analysis, the constraint

$$D_n D_m A = D_m D_n A \qquad (109)$$

can be satisfied even if "free" resonant terms are not included. If, on the other hand, the Lie-brackets of the two terms in the NF expansion do not vanish, than satisfaction of Eq. (109) requires the inclusion of "free" amplitudes.

As the algebra of vector fields, constructed from resonant monomials, is intimately related to the Lie-brackets of the same monomials [8], the conclusions may be also formulated in terms of that algebra. The results of [8] show that the sub-algebra with $l = 1$ is commutative. Namely,

$$\{\mathbf{S}_1^n, \mathbf{S}_1^m\} = 0 \quad \Rightarrow \quad [X_1^n, X_1^m] = 0 \qquad (110)$$

If the perturbation "samples" only the commutative sub-algebra, then the MMTS analysis can do without the "free" amplitudes. If the perturbation "samples" non-commutative components of the algebra then "free" amplitudes must be included appropriately.

## 5. Inability to determine dependence on $T_n$, $n \geq 2$: General case

Consider an $N$-dimensional system (the fast dependence on time can be always eliminated)

$$\frac{d}{dt}\mathbf{x} = \sum_{n \geq 1} \varepsilon^n \mathbf{F}_n \qquad (111)$$

In Eq. (111), $\mathbf{x}$ and $\mathbf{F}_n$ are row-vectors

$$\mathbf{x} = \left(x^{(1)}, x^{(2)}, \ldots, x^{(N)}\right), \qquad \mathbf{F}_n = \left(F_n^{(1)}, F_n^{(2)}, \ldots, F_n^{(N)}\right) \qquad (112)$$

(Superscripts denote the component of the vector). Expand $\mathbf{x}$ in a Near Identity Transformation (NIT)

$$\mathbf{x} = \sum_{n \geq 0} \varepsilon^n \mathbf{x}_n \qquad (113)$$

Through $O(\varepsilon^n)$, the MMTS expansion yields the following generic system of solvability conditions:

$$\begin{pmatrix} D_1 & & & & & \\ D_2 & D_1 & & & 0 & \\ D_3 & D_2 & D_1 & & & \\ . & . & . & . & & \\ . & . & . & . & . & \\ D_n & D_{n-1} & ... & ... & ... & D_1 \end{pmatrix} \begin{pmatrix} \mathbf{x}_0 \\ \tilde{\mathbf{x}}_1 \\ \tilde{\mathbf{x}}_2 \\ . \\ . \\ \tilde{\mathbf{x}}_{n-1} \end{pmatrix} =$$

$$\mathbf{S}(\mathbf{x}_0) + \begin{pmatrix} 0 \\ 0 \\ \mathbf{N}_2(\mathbf{x}_0, \tilde{\mathbf{x}}_1) \\ \mathbf{N}_3(\mathbf{x}_0, \tilde{\mathbf{x}}_1, \tilde{\mathbf{x}}_2) \\ . \\ . \\ \mathbf{N}_n(\mathbf{x}_0, \tilde{\mathbf{x}}_1, \tilde{\mathbf{x}}_2, \mathbf{K}, \tilde{\mathbf{x}}_{n-2}) \end{pmatrix} + \left\{ \nabla_{\mathbf{x}_0} \mathbf{F}_1(\mathbf{x}_0) \begin{pmatrix} 0 \\ \tilde{\mathbf{x}}_1 \\ \tilde{\mathbf{x}}_2 \\ \tilde{\mathbf{x}}_3 \\ . \\ . \\ \tilde{\mathbf{x}}_{n-1} \end{pmatrix} \right\}_R$$

(114)

In Eq. (114), $\mathbf{S}(\mathbf{x}_0)$ is a matrix of resonant terms that emerge in the analysis, and depend only on the components of $\mathbf{x}_0$. The free amplitude, which appears in the *k*'th order term ($1 \le k \le n-1$) in the NIT, is denoted by $\tilde{\mathbf{x}}_k$. In any order, $k \ge 3$, $\mathbf{N}_k(\mathbf{x}_0, \tilde{\mathbf{x}}_1, ..., \tilde{\mathbf{x}}_{k-2})$ is a matrix of resonant terms that is at least linear in each of the free amplitudes, $\tilde{\mathbf{x}}_l$, $1 \le l \le k - 2$. Each free amplitude, $\tilde{\mathbf{x}}_{k-1}$, $2 \le k \le n$, appears for the *first time* in the *k*'th-order solvability condition as a linear term [last term in Eq. (114)]. The index *R* in this last term denotes the fact that only the resonant contribution is to be included.

Eq. (114) constitutes a system of linear first-order differential equations for the $T_1$ dependence of all the amplitudes. It can be solved successively, starting from first order. The dependence on the higher time scales is not determined by the equations. Rather, it has to be provided either by the initial data at, say, $T_1 = 0$, or by other requirements. If these are appropriately smooth, then the commutativity of derivatives with respect to different time scales is a consequence of the equations, and does not induce new independent constraints. It becomes a convenient tool for finding the "free" amplitudes in complicated systems. Only if Lie brackets of all resonant terms in $\mathbf{S}(\mathbf{x}_0)$ vanish, can one set all the free amplitudes to zero, and Eq. (114) then yields the dependence of $\mathbf{x}_0$ on *all* time scales.

## 6. **Concluding remarks**

In the higher-order MMTS analysis of most dynamical systems, the "free" amplitudes (solutions of the homogeneous equation) that emerge in every order in the NIT expansion, Eq. (113) play an important role far beyond the fulfilment of the initial conditions. Their inclusion is essential for the internal consistency of the expansion, specifically, consistency among solvability conditions that pertain to different orders. The expansion may yield trivial incorrect results if consistency is violated.

Only in special dynamical systems "free" amplitudes may be dropped. Even in the case of harmonic oscillatory systems that are perturbed by nonlinear perturbations, choice of the "free" amplitudes is not restricted only for a single oscillator with an energy conserving perturbation and an autonomous equation of motion. (A time-dependent interaction is equivalent to the introduction of an additional degree of freedom.) The latter system belongs to the limited class of dynamical systems for which the solvability conditions yield the dependence of the solution on *all* time scales: Systems for which the Lie brackets of pairs of resonant terms in $\mathbf{S}(\mathbf{x}_0)$ in Eq. (114) vanish [8] for all *m* and *n*,

$$\{\mathbf{S}_m(\mathbf{x}_0), \mathbf{S}_n(\mathbf{x}_0)\} = 0 \tag{115}$$

For given values of *m* and *n*, Eq. (115) ensures that the corresponding solvability conditions are mutually consistent even when the free amplitudes are set to zero. Namely, they do not violate the trivial requirement that derivatives with respect to the time scales, $T_m$ and $T_n$, commute:

$$D_m D_n \mathbf{x}_0 = D_n D_m \mathbf{x}_0 \tag{116}$$

Even simple systems, such as single harmonic oscillators with dissipative perturbations (see Section 2) and coupled harmonic oscillators with an *energy conserving* perturbation (see Section 3) are not in this class, because the Lie brackets of pure $(A, A^*)$ resonant terms do not vanish in general. Hence, the free amplitudes must be included. Clearly, when higher-dimensional systems are analyzed through orders beyond first, then the "free" terms must be retained, so as to ensure a consistent expansion.

Owing to the triangular structure of the differential operator in Eq. (114), the MMTS scheme cannot determine the dependence of the amplitudes on *all* time scales. In $O(\varepsilon^k)$, the l.h.s. of Eq. (114) is

$$D_k \mathbf{x}_0 + D_{k-1}\tilde{\mathbf{x}}_1 + D_{k-2}\tilde{\mathbf{x}}_2 + ... + D_1\tilde{\mathbf{x}}_{k-1}$$

This sum cannot distinguish among the contributions of different amplitudes to the $O(\varepsilon^k)$ component in the time dependence of the solution. Rather, it can only determine the $T_1$ dependence of $\tilde{\mathbf{x}}_{k-1}$, given the dependence of all lower-order amplitudes on $T_1$. To obtain the dependence of the amplitudes on higher time-scales, one must employ initial data at, say, $T_1 = 0$, or resort to one's intuition.

Satisfaction of the consistency requirements is equally important in the perturbative analysis of spatially extended dynamical systems, where, often, higher orders in the MMTS analysis are involved in the derivation of *amplitude equations*. As a PDE may be converted into an infinite system of ODE's for the time dependence of the Fourier coefficients obtained in a Fourier expansion of the solution in the spatial coordinates, the problem encountered in the examples of systems of two degrees of freedom may emerge. That the free resonant terms do play an important role is implicit in [13], where the general formalism for the derivation of the NLS equation is presented, with special attention paid to consistent satisfaction of higher-order solvability conditions. However, the analysis of perturbed PDE's is far more complicated than that of ODE's, so that the origin of the consistency problem may become obscured.

In the case of operator equations, the limitations of the MMTS are encountered in the analysis of even the simplest dynamical systems. The Lie brackets then become commutators, which, in general, do not vanish. Internal consistency of the set of solvability conditions is lost if the "free" operators (solutions of the homogeneous equation) are omitted. This has been recently shown to hold even in the case of the Quantum Duffing oscillator [14], which, in its classical version, does not suffer from the consistency problem.

As a final comment we note that, in the method of Normal Forms [9], there are no limitations on the free terms that emerge in every order of the expansion [11, 15–17].

Acknowledgments: Useful discussions with G. Burde, Y. Kodama, I. Rubinstein, L. Segel and B. Zaltzman are acknowledged.